\documentclass{ws-ijmpd}
\usepackage[super,compress]{cite}

\usepackage{graphicx}
\usepackage{dcolumn}
\usepackage{bm}
\usepackage{amsmath}
\usepackage{amsfonts,amssymb}
\usepackage{color}
\usepackage{enumerate}
\usepackage{float}
\usepackage{placeins}
\usepackage{subfig}
\usepackage[T1]{fontenc}
\usepackage[utf8]{inputenc}
\usepackage[polutonikogreek,english]{babel}
\begin{document}

\markboth{C.ORTIZ}
{SURFACE TENSION}
%
\catchline{}{}{}{}{}
%

\title{SURFACE TENSION: ACCELERATED EXPANSION COINCIDENCE PROBLEM  \& HUBBLE TENSION}

\author{C. ORTIZ}
\address{Unidad Acad\'emica de F\'isica, Universidad Aut\'onoma de Zacatecas,\\ Calzada Solidaridad esquina con Paseo a la Bufa S/N C.P. 98060, Zacatecas, M\'exico \\ ortizgca@fisica.uaz.edu.mx}

\maketitle

\begin{abstract}
 In this paper we give   a physical explanation to the accelerated expansion of the Universe, alleviating the tension between the discrepancy of Hubble constant measurements. By the Euler–Cauchy stress principle, we identify a controversy on the lack of consideration of the surface forces contemplated in the study of the expansion of the Universe. We distinguish  a novel effect that modifies the space-time fabric by means of the energy conservation equation.  The resulting dynamical equations from the proposed hypothesis are contrasted to several  testable astrophysical predictions.  This paper also explains why we have not found any particle or fluid responsible for dark energy and  clarifies  the  Cosmological Coincidence Problem. These explanations are achieved without assuming the existence of exotic matter of unphysical meaning or having to modify the Einstein's Field Equations.

\end{abstract}

\keywords{General Relativity; Cosmology; Dark Energy; Hubble Tension.}

\ccode{PACS numbers:98.80.Jk; 98.80.Es; 68.03.Cd.}


\section{Introduction}\label{Int}

The Einstein Field Equations (EFE), describe the fundamental interactions of gravitation as a result of space-time being curved by means of the mass and energy inside this space-time. In this sense the metric and the stress-energy tensor determine the system, and  the dynamics of it can be obtained via the variation principle of the Einstein Hilbert (EH) action plus the matter  action field.

In 1917, Einstein  realized  that the dynamics of his new theory predicted a non permanent universe, simply  because all  the matter attracts gravitationally. Consequently, influenced by the  belief, at the time, that the Universe was static and everlasting, he decided to modify  his theory,  including a constant parameter, called the Cosmological Constant (CC). Short after, in 1922, Alexander  Friedmann \cite{Friedmann}, sets bases for the theoretical model for an expanding Universe. The accelerated expansion of the Universe was confirmed with observational evidence by   George Lamaître \cite{Lematitre} in 1927,  and by  Edwin Hubble \cite{Hubble}, in 1929;  establishing the correlation  between   the redshift and the   distance  from its sources.

To the moment, the   Lambda Cold Dark Matter ($\Lambda$CDM) cosmological  model is  the best candidate to explain  the accelerated expansion of the Universe. The power spectrum of the flat $\Lambda$CDM model  is the  one that best fits the cosmological observations, like Cosmic Microwave Background Radiation (CMB) \cite{penziaswilson,dicke}, Baryonic Acoustic Oscillations (BAO) \cite{eisenstein,percival} and large scale structure formations \cite{spergelwmap,spergelwmap2,komatsuwmap,liddlestructure}, among others \cite{stoughtonsloan,abazajiansloan,tegmarksdsswmap}. Most of the parameters of the model are well understood, except  the CC, the  one parameter that rules the  accelerated evolution of the Universe \cite{Ade:2015xua}. 

There is a strong disagreement between the predicted dark energy density and the observed one, which is of the order of magnitude of the current matter energy density. This problem is known as the Cosmological Constant Problem  \cite{RevModPhys}.

    Under the standard cosmological model $\Lambda$CDM, recent measurements  of the expansion rate of the Universe, at low redshifts  \cite{Riess:2019}, appear to be in disagreement with the predictions for observations of the   CMB \cite{planck:2018}, this disagreement is   known as the Hubble's Tension \cite{Freedman:2017yms}.

 Even more so, the fact that the ratio of the baryonic energy density to  dark energy density at present time is of the order of one in the $\Lambda$CDM model, makes us look up  for a physical explanation for such a remarkable coincidence, the Cosmological Coincidence Problem \cite{CCP}.
   
Scientists have been proposing many alternative models and theories, in order to explain the expansion phenomenon, the Coincidence Problem and  Hubble's Tension. Basically,  such explanations can be sorted into two different kinds \cite{Joyce}: one is that the problem depends on the matter content, i.e. the stress-energy tensor, $T_{\mu \nu}$, and that there must be  some  extra energy fluid    that may fill the space-time, e.g., quintessence \cite{Phys}; another possibility states that the problem rests  on the geometric sector, implying that Einstein's theory could be wrong or incomplete e.g., Modified Gravity \cite{CLIF}.

In order to address these problems,  in this paper we consider  the surface forces of a homogeneous and isotropic cosmological system. Taking into account the surface  energy of the analyzed system,  a  new relativistic effect, never considered before, is derived, modifying  the system of equations that governs the evolution of the expansion of the Universe.   The derived contribution, alleviates the Hubble tension, being in agreement with the measurements of the expansion of the Universe  at late and early epochs.   The proposed model has the desired  futures of the  $\Lambda$CDM model; furthermore, it explains the  accelerated  expansion of the Universe \cite{perlmutter, riess, Ade:2015xua}, without adding extra exotic matter. 
                
 This paper is organized as follows:  In Sec. \ref{tensionc}, the Euler-Cauchy stress principle is applied to the thermodynamic equation, arriving to a density evolution equation. In  Sec. \ref{renatax}, the equation of the  dynamics of the Universe is stated, by taking into account both effects: the gravitational one, due to the inner region, plus a   general relativistic effect due to the  surface tension. In sec. \ref{cosmo} the derived equations of the evolution of the Universe are restated in terms of observable parameters. Sec. \ref{methodology}, presents a description of the methods and the data used to constrain  the cosmological parameters of the proposed  model.  Sec. \ref{results},  shows the best estimate of the constrained parameters and the fixed proposed model  is contrasted to different astrophysical observations, highlighting the advantages of the model against the standard cosmological model.   Finally, in  Sec. \ref{Dis}, we discuss our results.

\section{Surface Tension} \label{tensionc}
 The present section is devoted to introduce the hypothesis  of the surface tension, $ \boldmath \gamma_S$, and its implications in the  dynamics of a  perfect fluid  density.

Surface tension is associated with the nature of the chemical bonds of atoms (electromagnetic force) at the surface of an interface between different materials; nevertheless, the surface tension is related to   the Euler–Cauchy stress principle  which is a basic  concept of continuum mechanics \cite{fung2001classical}. This principle   states that upon any close surface (real or imaginary) that divides a body, the action of one part of the body on the other is equivalent to its external forces acting on it. For bodies in continuous media,  there are two types of external forces:
 \begin{itemize}
     \item Body forces, $(\mathbf f)$.
     \item Surface forces or stress. $(\mathbf F)$.
 \end{itemize}
Thus, the total force ${\displaystyle {\mathcal {F}}}$ applied to a body or to a portion of the body is the contribution of all the forces.

The interactions between pairs of atoms or molecules, are usually modeled with the Lennard-Jones potential. This is a good approximation for short interaction distances, since it is related to the Pauli exclusion principle and the Van der Waals force due to intermolecular forces. The gravitational potential, being   several orders of magnitude less  than the electromagnetic potential, is  neglected. 

We  know that the gravitational potential is the one that acts at large distances, so we  will work under the hypothesis that surface tension due to this potential is relevant at large scales. 

The required work to increase a given surface area due to the surface  tension is given by:

\begin{equation}\label{tension}
      dW_S= \gamma_S dA.
\end{equation}

Considering the first law of thermodynamics for a closed system 

\begin{equation}
      d(U)= \delta Q - \delta W,
\end{equation}
for a non covariant theory  the energy $U$ is the  integral of the energy density, $ \delta Q $ is the infinitesimal heat introduced to the system by its surroundings, and   $ \delta W$ is the work done by the system to its surroundings.

If we only take  into account the surface energy of the system   (\ref{tension}), for a given total energy $U$:
\begin{equation}
      d(U)= -\gamma_S dA,
\end{equation}
where the total energy of the system, $U$, is given by the mass-energy relation $U=mc^2$, and based on the cosmological principle we  assume spherical symmetry, then,
\begin{equation}
      d(\rho V )= -\frac{\gamma_S}{c^2} 8\pi R dR,
\end{equation}
integrating from $R=0$ to $R=r$
\begin{equation}
        \rho V-\rho_0 V_0= -\frac{\gamma_S}{c^2} 4\pi r^2.
\end{equation}
The explicit  form of the surface tension term, $\gamma_S$, depends on the symmetries of the problem and the external forces.  In our set-up  model, we have spherical symmetry; so  in order to simplify the calculations we choose the surface tension of a bubble,
\begin{equation}
\gamma_S=\frac{F}{4r}, 
\end{equation}
the limitations of the above consideration are set by the cosmological framework, in this sense, the distance is limited by de Hubble radius.

By taking  into account that the gravitational force is the only one  that acts at long distances, it yields 
\begin{equation}
        \rho V=\rho_0 V_0 +\frac{Gmm_0}{r_0^2 c^2}\pi r.
\end{equation}

Considering that the  gravitational interaction is given by a Planck mass, $m_0$, over a Planck radius, $r_0$, the following relation is fulfilled 
\begin{equation}
\frac{Gm_0}{r_0 c^2}=1.
\end{equation}
By aid of the previous relation, we arrive to the evolution  density equation:

\begin{equation}\label{densidad}
 \rho =\rho_0 \frac{V_0}{V} +\rho_0\pi \frac{r}{r_0}.
\end{equation}
The resulting  density equation takes into account the matter content inside the system and the surface tension of the given system. It is worth noting that the initial density $\rho_0$ in equation (\ref{densidad}) is the same for both RHS terms, since it is a homogeneous system.
 
So as  to  consider some extra different density  fluids that do not interact among them, the superposition principle is valid, so the resulting density equation would be a linear combination of the different fluids densities.

Based on the Euler-Cauchy stress principle,  equation (\ref{densidad}) is true for any spherical surface chosen in the cosmological domain.


Being all the matter content analyzed, we proceed to  contrast the effects of the surface tension on the dynamics of the Universe.
\section{Dynamics of the Universe}\label{renatax}
Here  we present the equation of the dynamics of a flat homogeneous and isotropic   Universe, due to the barionic and radiation matter contents, plus the energy density due to the surface tension.

Departing from the variation of the action of the flat Friedmann-Robertson-Walker metric (FRW), we get to the well known Friedmann equation, 
\begin{equation}\label{fi}
3\left(\frac{\dot{a}}{a}\right)^2= 8\pi G \rho,
\end{equation}
where $\rho$ is the  total density of the  system.

When we introduce  the  density due to the matter content and the  surface tension (\ref{densidad}), we must note that the proposed model framework is in a generally covariant one, so  the total energy of the system is given by the  Tolman-Komar energy relation \cite{tolman}, which includes the perfect fluid pressure contributions.  

The  modified  Friedmann equation (\ref{fi}), reads as,

\begin{equation}\label{tab}
\left(\frac{\dot {a}}{a}\right)^2=\frac{8 \pi G}{3}\rho_{0m}\left(\frac{a}{a_0}\right)^{-3}+\frac{8 \pi G}{3}\rho_{0r}\left(\frac{a}{a_0}\right)^{-4} +\frac{8 \pi G}{3}\rho_{0m}\pi \left(\frac{a}{a_0}\right),
\end{equation}
the LHS of the equation provides  the kinetic information; the first and second  RHS  equations give the information of the  different fluids,  dust and radiation; while the last term  relates to  the surface tension.  We must note that a  time  parametrization  of the distance, $ {\bf{r}}= a(t){\bf\hat{r}}$, was made.

The resulting  acceleration equation is given by,
\begin{eqnarray}\label{fine}
\frac{\ddot a}{a}&=&-\frac{4 \pi G}{3}\rho_{0m}\left(\frac{a}{a_0}\right)^{-3} -\frac{8 \pi G}{3}\rho_{0r}\left(\frac{a}{a_0}\right)^{-4}+4 \pi G \rho_{0m} \pi \left(\frac{a}{a_0} \right),
\end{eqnarray}
where the first two  terms on the RHS of equations (\ref{tab}) and (\ref{fine}) gather all the different fluids inside the inner region, including possibly, the dark matter; while   the last term    on   both equations, gather all the different kinds of matter in the boundary region.

It is worth noting, that the positive sign on the  last term of the  acceleration equation (\ref{fine}),  is  responsible for the accelerated expansion of the Universe; this acceleration is due to the surface energy of the considered system. We also notice that this  term  is time dependent.

 Under these circumstances, the surface energy  acts  much like the cosmological constant. This term   gives  the information on how the surface tension  stretches out the space-time fabric,
which  explains  why there is no such a dark energy fluid or particle. 

These equations  resemble the Friedmann Equations of a $\Lambda$CDM  model, with the advantage of one less parameter  to be fixed in the model, since $\rho_0$ is obtained by the system conditions.

The present model satisfies the Dominant Energy Condition (DEC) $\rho\pm P\geq 0$, since $\rho>0$ as seen on equation (\ref{densidad}), and we only consider dust and radiation matter content. Under this hypothesis, the accelerated expansion of the Universe is the result of the surface tension, rather than the consequence of a negative pressure.

 
\section{Model Parametrization}\label{cosmo}

The information that our Universe is at an accelerated expansion epoch, comes from the observation of the redshifts of the frequency of  light emitted by distant sources.

In order to contrast our proposed model with the cosmological observations, we rewrite the Friedmann equation (\ref{tab}) and (\ref{fine}) in terms of the redshift parameter.

The scale factor $a(t)$ is related to the  redshift,  $z$, by the following equation, 
\begin{equation}
\frac{a(t)}{a(t_0)}=\frac {1}{1+z}.
\end{equation}

 The Friedmann equation in terms of the redshift becomes,
\begin{equation}\label{vel}
H(z)^2=\frac{8 \pi G}{3}\rho_{0m}\left(1+z\right)^{3}+\frac{8 \pi G}{3}\rho_{0r}\left(1+z\right)^{4} +\frac{8 \pi G}{3}\rho_{0m}\pi \left(1+z\right)^{-1}.
\end{equation}
The critical density,  $\rho_c$, is defined as:
\begin{equation}
    \rho_c \equiv \frac{3H^2}{8\pi G},
\end{equation}
and the density parameters are  given by  the relation $\Omega_{i}\equiv 8 \pi G \rho_{0i}/3 H^2$.
 
If we divide equation (\ref{vel}) by  $H(z)^2$ at $z=0$, we get a critical density relation
\begin{equation}\label{fried}
  \Omega_{0m}  + \Omega_{0r} + \Omega_{0m}\pi=1.
\end{equation}

  If we consider that at present time, the radiation density component, is several orders of magnitude less than the matter density parameter, we have that
 
 \begin{equation}
  \Omega_{0m} =\frac{1- \Omega_{0r} }{1+\pi}\approx 0.24145.
\end{equation}
 This relation simplifies the constriction of the model. 
 
   We must note  that the first term on the  RHS of the critical density relation (\ref{fried}),  is the matter energy density term, $\Omega_{0m}$, and that this term  is  of  the same order of magnitude as the third term,  the energy density of the surface tension,  $\Omega_{0m} \pi$,  which is responsible of the accelerated expansion of the Universe. The nature of this last term gives an  explanation to the Cosmological Coincidence Problem \cite{RevModPhys}.

In order to constrain the free parameters of our model,  we cast the Hubble  parameter as a normalized Hubble parameter $E(z)$, written in terms of the density parameters and the redshift. 
\begin{equation}
E^2(z)=\left(\frac{H(z)}{H_0}\right)^2,
\end{equation}
so the resulting normalized Friedmann Equation for the ST model reads:
 \begin{equation}\label{tab3}
E^2(z)=\Omega_{0m} (1+z)^3 + \Omega_{0r} (1+z)^4+ \Omega_{0m}\pi (1+z)^{-1}.
\end{equation}

If we write $\Omega_{x0}=\Omega_{m0}\pi$, it would  correspond to   phantom Dark Energy in  a $w$CDM model, with  $w=-4/3$.  It is known that this theories exhibit some pathologies e.g. it violates the DEC, leading to vacuum instabilities.  It is worth pointing out that  the proposed  model indeed satisfies the DEC, as it was mentioned on the previous section.


With aid of the normalized Friedmann equation (\ref{tab3}), the deceleration parameter (\ref{des1}) and  the redshift of different objects, we can calibrate the proposed model.

\section{Data and Methodology}\label{methodology}

We will constrain the free parameters of the model, by minimizing the merit of function $\log \mathcal{L} \sim$ $\chi^2$. The model will be tested with different observational data sets: Observational Hubble Data (OHD) and Type Ia Supernovae (SNIa) distance modulus.

\subsection{Observational Hubble Data}\label{Hubble}

 We calculate the optimal model parameter, $H_0$,   by minimizing the function of merit, 
\begin{equation}\label{chi}
\chi_{H}^2=\sum_{i=1}^{N_H} \left(\frac{H_{th}(z_i, \Theta)-H_{obs}(z_i)}{\sigma_{obs}^{i}} \right)^2.
\end{equation}
Where $H_{th}$ is  the value of the Hubble parameter of the theoretical model with parameter space $\Theta(h)$; $H_{obs}$  and  $\sigma_{obs}^{i}$ are the observational Hubble parameters from a given sample and its correspondent uncertainty.

 The sample  consists of  $N_H=34$ $H(z)$ measurements  in the redshift  range  $0.09 < z< 2.36$. The measurements  comes from Baryon Acoustic Oscillations (BAO) \cite{Beutler_2011,DR12,DR14BOSS,DR14234,DR14Lalpha235} 
 and Cosmic Chronometers  \cite{loeb}.

Once the model is constrained, we will compare the proposed model to the Standard Cosmological Model using  the Beyesian Information Criterion (BIC) \cite{BIC} defined as:
\begin{equation}
    BIC= \chi_{min}^2+k \ln{N},
\end{equation}
where $\chi_{min}^2$ is log-likelihood of the model, k is the number of free parameters of the optimized model and $N$ is  the number of data samples.  This criterion  gives us a quantitative value to select among several models. The model with the lower BIC,  is  the  most favoured.

 
\subsection{SNIa Supernovae}\label{SNI}

 The  data from SNIa observations is usually released as a distance modulus $\mu$. This cosmological parameter allows us to constrain or contrast a cosmological model.

 From the apparent magnitude $m_b$ which is related to the luminosity distance

\begin{equation}
    d_L(z)=\frac{(1-z)}{H_0}\int_0^z\frac{dz'}{E(z')},
\end{equation}
 and with aid of the absolute magnitude $M$, we can calculate the  distance modulus

\begin{equation}
   \mu=m_b-M=5\log_{10}\left(\frac{d_L}{Mpc}\right)+25.
\end{equation}

This relation allows to contrast our theoretical model to the observations, by minimizing the function of merit,
\begin{equation}\label{chidm}
\chi_{\mu}^2=\sum_{i=1}^{N_{\mu}} \left(\frac{\mu_{th}(z_i, \Theta)-\mu_{obs}(z_i)}{\sigma_{obs}^{i}} \right)^2.
\end{equation}

The observational data compilation    for the distance modulus will be  the Pantheon Type Ia catalog \cite{pantheon}, which consists of $N_{\mu}=1048$ SNe data samples, that  includes observations up to redshift $z=2.26$. It will  be assumed  a nominal value of $M=-19.3$\cite{moduluscorrection}.

\section{Analysis and Results}\label{results}

In the present section we report the  best fit parameters for the surface tension model. We  present some cosmological  parameters and features of the model,  including comparative graphics derived from  the obtained results.

 The congruence of the proposed model is demonstrated under the different performed analyses. We  also show how  the proposed model  addresses the Hubble tension problem and the age of the Universe, as well as the behaviour of the deceleration parameter.

\subsection{Hubble Expansion }\label{hubble}

As  shown on the study from the Planck collaboration 2018 \cite{planck:2018}, there is a considerable discrepancy on the measurements of the expansion of the Universe at  different redshifts, and the prediction of the expansion from the CMB analysis. This is true for the base   $\Lambda$CDM and some simple model extensions analyzed in it.

The disagreement on the measurement of the expansion of the Universe \cite{planck:2018}, can be attributed to  two  possible reasons.  The first one is that the discrepancy is due to an error on the measurements; the second one could be a problem with the current model of parametrization of our Universe, the $\Lambda$CDM model. 
As  emphasized by Riess et al., \cite{Riess:2019} the independent realized tests   rule out the possibility that the discrepancy is due to errors on the measurements, and   points out  that the  problem depends on the employed model, the  $\Lambda$CDM model, or in a cosmological feature beyond it.

Here we compare the proposed model to the latest measurements and  contrast it to the standard cosmological model. The results are shown in the following table,

\begin{table}[h]
\centering
\begin{tabular}{|l|l|l|l|}
\hline
Model &  $H_0  $  Km s$^{-1}Mpc^{-1}$ & Matter density,  $ \Omega_{m} $&    BIC \\ \hline
ST &$74.63_{-2.7}^{+3.2} $ & $0.24145 $&  26.73  \\ \hline
 $\Lambda$CDM & $69.51_{-3.4}^{+3.9} $ & $0.26152$ & 37.17 \\ \hline
\end{tabular}
\caption{Best estimate of the constrained   parameters for ST model and $\Lambda$CDM model.}
\label{table}
\end{table}

The difference on the BIC parameter, $\Delta$BIC$=10.45$, is a very strong evidence in favor of  our proposed model.

\begin{figure}[H]
\centering
\includegraphics[scale=0.7]{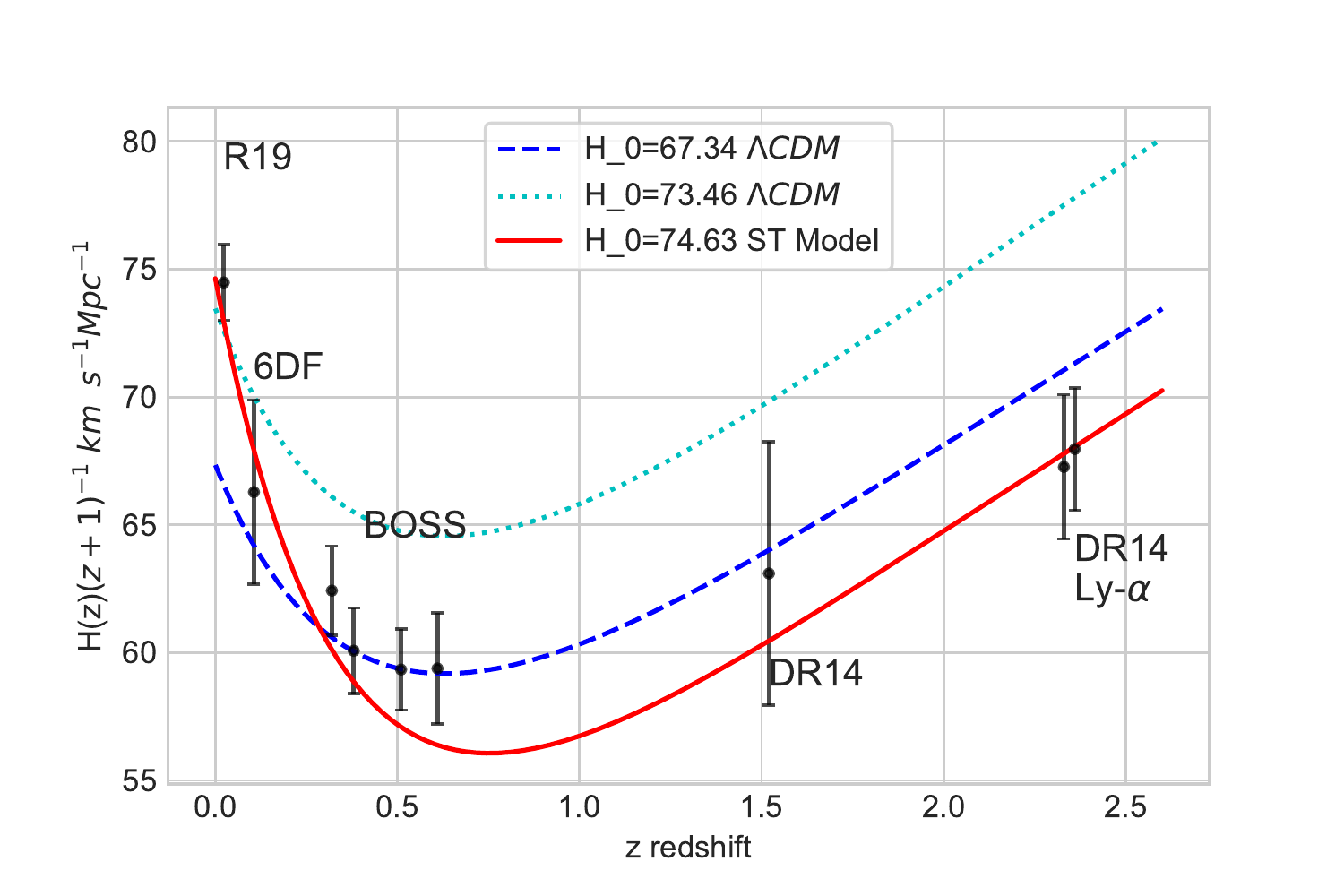} \\
\caption{ Comoving Hubble parameter, H(z)$/(z+1)$,  as a function of redshift between  $0<z<2.5$. The red line represents the proposed model with a Hubble parameter of  $H_0=74.63$ Km$s^{-1}$Mpc $^{-1}$,  baryonic density $\Omega_m =0.24145$, and radiation density of $\Omega_r =7.08 \times 10^{-5} $.}
\label{fig2}
\end{figure}

\begin{figure}[htbp]
\centering
\includegraphics[scale=0.7]{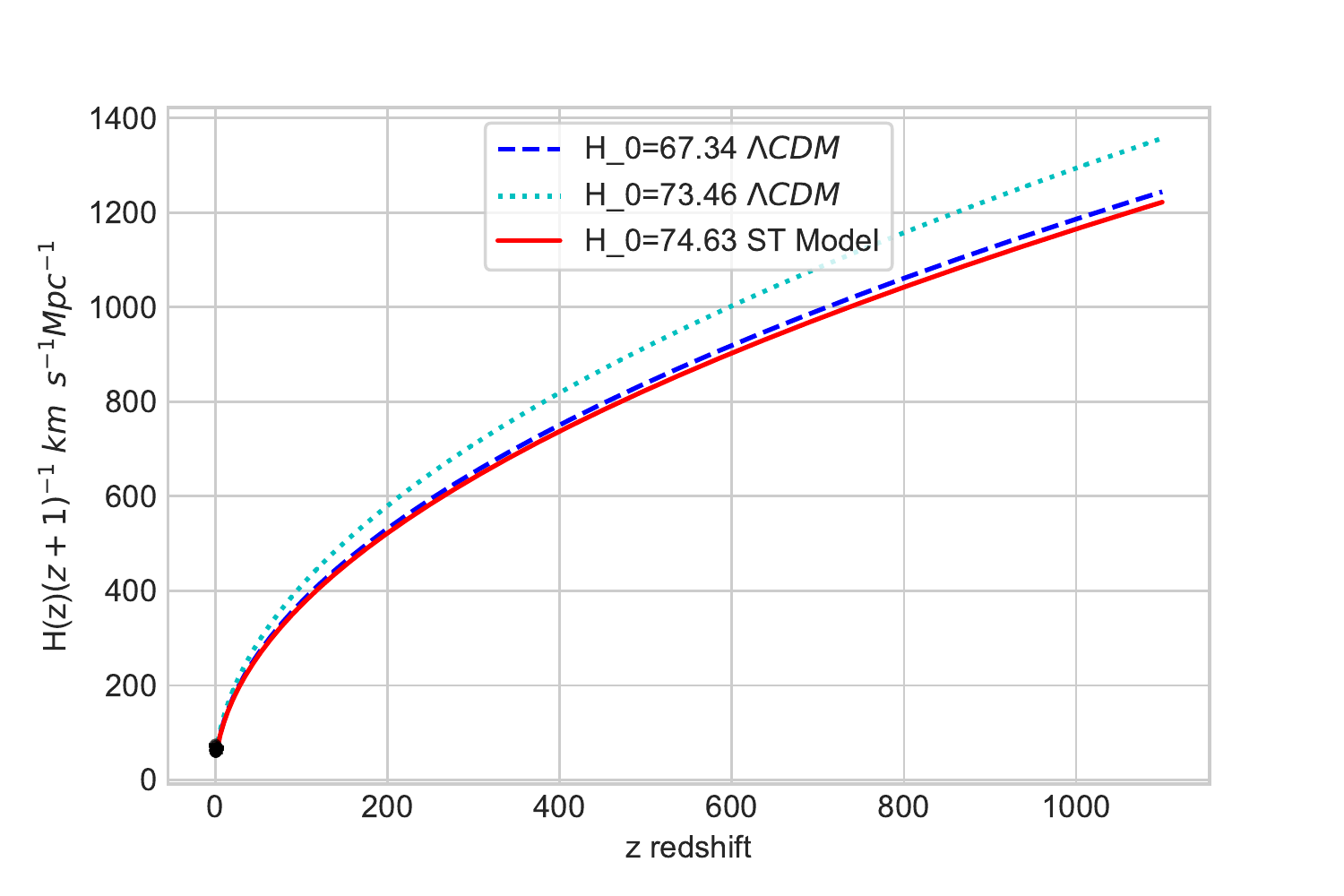} \\
\caption{ Comoving Hubble parameter, H(z)$/(z+1)$,  as a function of redshift, between $0<z<1100$.} 
\label{fig3}
\end{figure}
\FloatBarrier
If we contrast the  $\Lambda$CDM to the our model, it can be seen that, the former is in good agreement with measurements of the $H(z)$ for the 6DF Galaxy measurements  \cite{Beutler_2011},  BOSS DR12 \cite{DR12}, DR14 quasars \cite{DR14BOSS}, DR14 correlations of Ly-$\alpha$ \cite{DR14234}  and Ly-$\alpha$ cross-correlations  \cite{DR14Lalpha235}, all of them  with  redshifts between $0.3$-$2.5$; but there is a huge discrepancy    when we contrast it  to  low redshifts  R19 measurements \cite{Riess:2019}.   Our proposed model is in excellent accordance    with both redshift reneges, as it can be seen on figures (\ref{fig2}) and   (\ref{fig3}).

Under our stated hypothesis, that the surface tension is  responsible for the accelerated expansion of the Universe,  we can conclude that there is no such Hubble's tension; the problem is  not  in the measurements data technique, but in the employed  cosmological model.   
It is important to emphasize that in order to select the optimal model parameter, the data from Riess et al.\cite{Riess:2019} was not included. In spite of the lack of inclusion of this data, the proposed model predicts those observations.
\subsection{Luminosity Distance}\label{supernovae}
Type Ia Supernovae  homogeneity and its  well determined absolute magnitude as a function of its light curves allow us to  indirectly determine extragalactic distances with good accuracy. The best fit parameters of the model  is $H_0=74.94 $. This value is practically the same that we obtained  via OHD  on the previous subsection (\ref{hubble}).

In order to plot the distance modulus for the model, $\mu$, we use the best fit    and  contrast it to  the observational data from  the Pantheon compilation \cite{pantheon}.
\begin{figure}[htbp]
\centering
\includegraphics[scale=0.7]{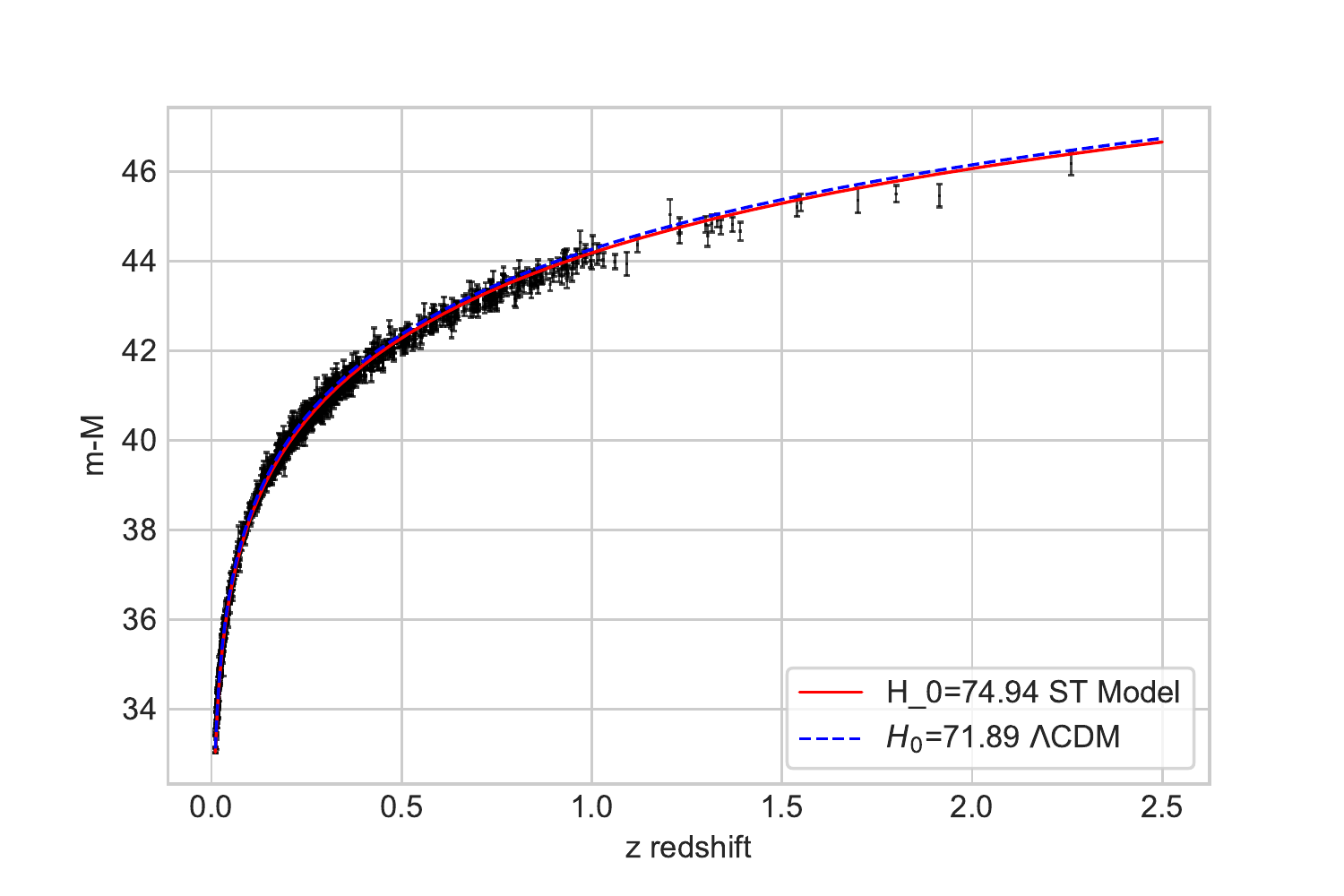} \\
\caption{Hubble  diagram for the Pantheon compilation, Distance Modulus as a function of redshift. } 
\label{lux}
\end{figure}
\FloatBarrier

\begin{figure}[htbp]
\centering
\includegraphics[scale=0.7]{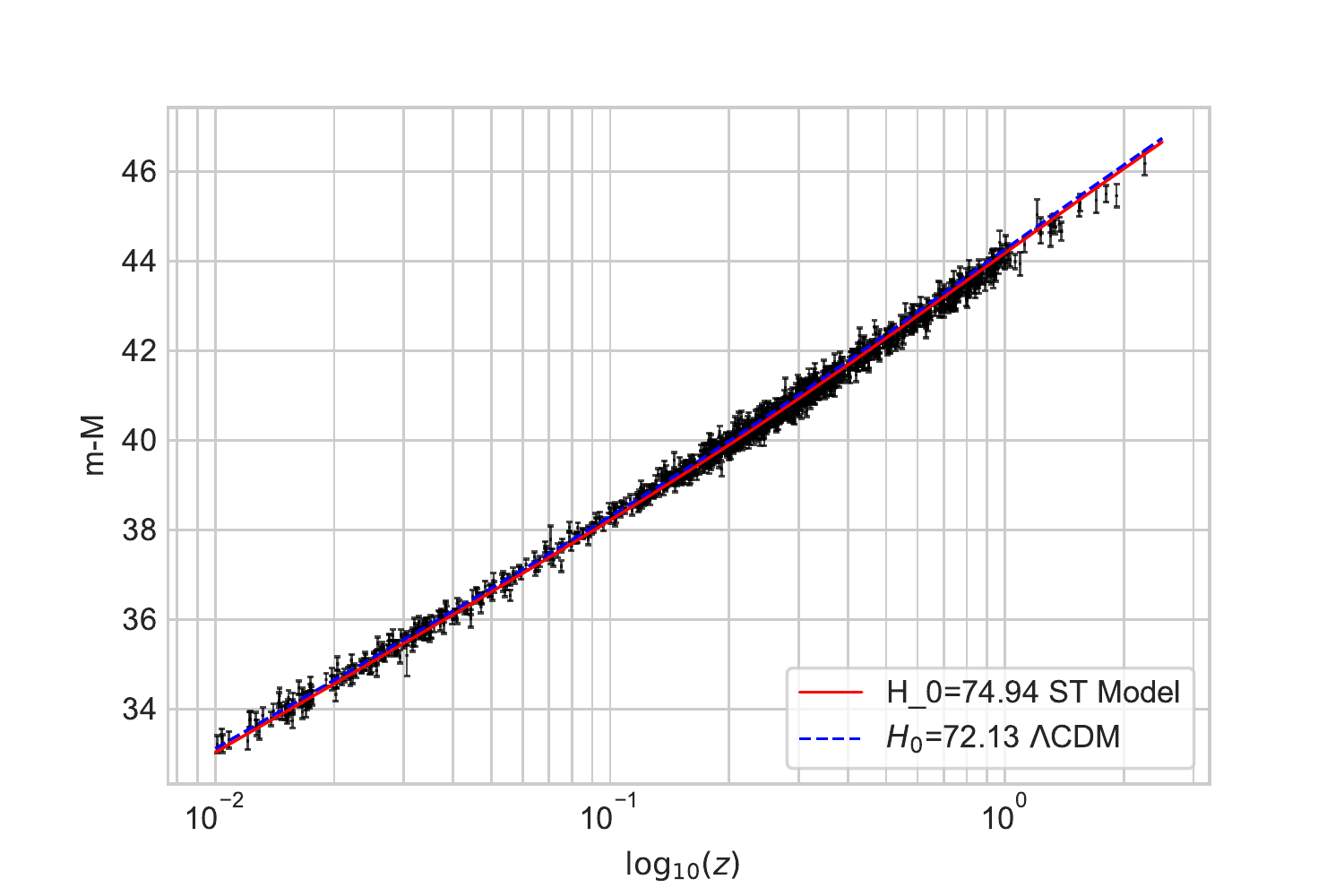} 
\caption{ Hubble  diagram for the Pantheon
compilation, Distance Modulus as a function of redshift, in logarithmic scale.} \label{loglux}
\end{figure}
\FloatBarrier

The plot of the distance modulus in a  logarithmic scale, presented on figure (\ref{loglux}), gives a better perspective at low redshifts.

We can appreciate  from the  distance modulus plots, Figures (\ref{lux}) and (\ref{loglux}), the congruence of the proposed model and the astronomical data  from the Pantheon catalog.  \cite{pantheon}

We must point out, that the obtained value is in agreement with the Hubble parameter $H_0=74.03 \pm$ 1.42  Km s$^{-1}Mpc^{-1}$  predicted by Riess et al. \cite{Riess:2019}

\subsection{Age of the Universe }\label{age}
 There are several ways to  estimate the age of the Universe. For a given cosmological model, the age can be estimated   with the aid of the Friedmann equation,  its relationship is given by:  

\begin{equation}
    t_0=\frac{1}{H_0}\int_0^\infty \frac{dz}{(1+z)H(z)}.
\end{equation}

It is known that the $\Lambda$CDM is the most consistent model with the cosmological observations, the estimated age  of the Universe is $13.801 \pm 0.024$ Gyr \cite{planck:2018}.

The age of the Universe for the best fit surface tension model, is $13.93$ Gyr., slightly more than the predicted by the standard cosmological model, but certainly  not in tension.

\subsection{Deceleration Parameter}\label{decelera}
 It is  a fact that the Universe is at an accelerated expansion epoch\cite{Lematitre,Hubble}.  To measure the cosmic acceleration of the expansion of the Universe we use the deceleration parameter, that is given by the relation, 
\begin{equation}
q(z)=-\frac{\ddot{a}}{a}\left(H_0 E(z)\right)^{-2},
\end{equation}
in terms of the redshift and $E(z)$,

\begin{equation}
  q(z)=-1+\frac{(1+z)}{E(z)}\left(\frac{d E(z)}{dz}\right),
\end{equation}
 by introducing equation (\ref{tab3}) we arrive to the deceleration parameter for our proposed model,

\begin{equation}\label{des1}
  q(z)=\frac{1}{2E(z)^2}\left(\Omega_{0m} (1+z)^3 + 2\Omega_{0r} (1+z)^4- 3\Omega_{0m}\pi (1+z)^{-1} \right).
\end{equation}

It can be seen that the only responsible of the accelerated expansion of the Universe  is the third term on the last equation, $3\Omega_{0m}\pi (1+z)^{-1}$, this term comes  from considering the surface tension, and it depends on the matter density of the system.

\begin{figure}[htbp]
\centering
\includegraphics[scale=0.7]{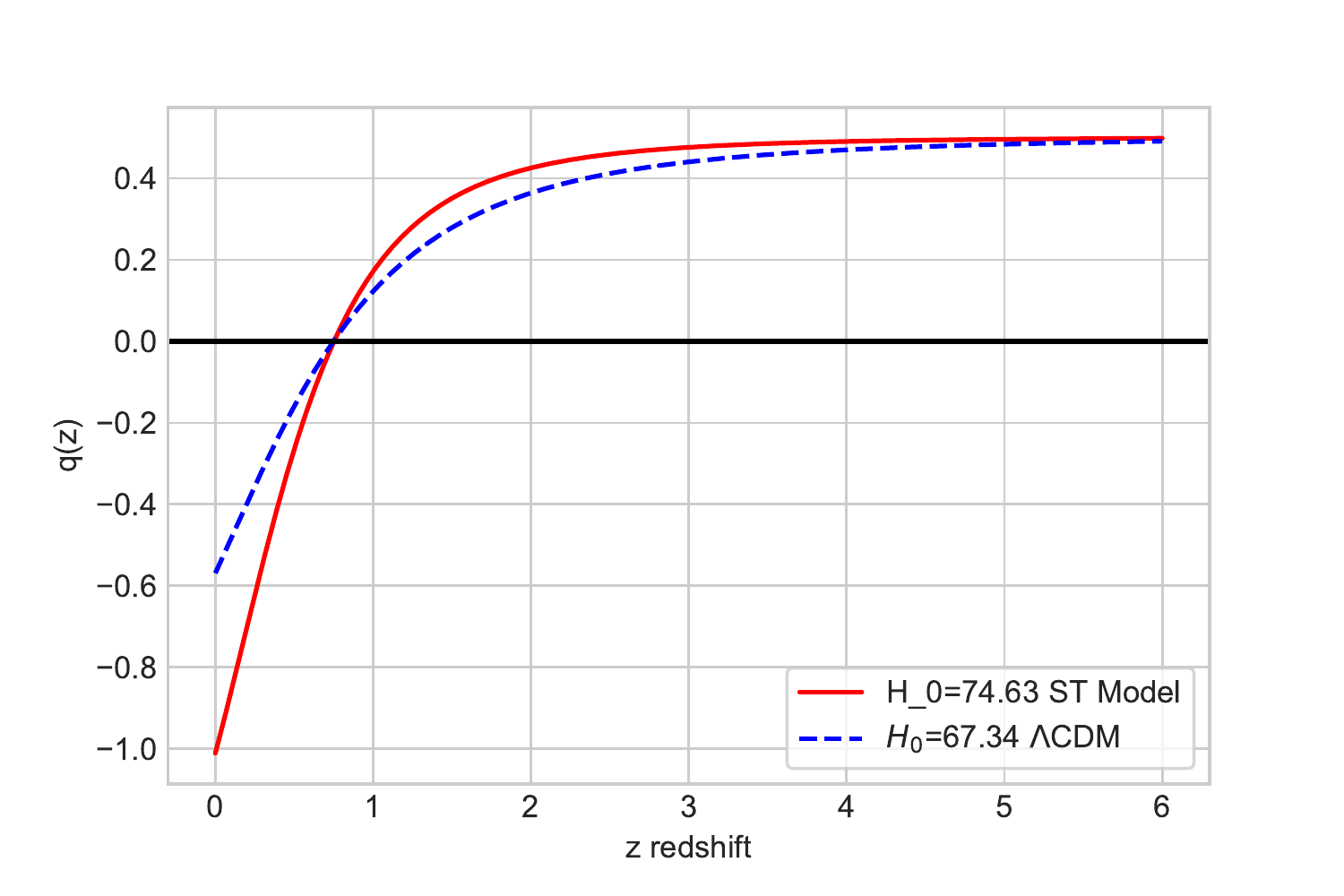} \\
\caption{Evolution of the deceleration parameter in terms of the redshift for the $\Lambda$CDM  model and the proposed model. \label{deceleration}} 
\end{figure}

The best   present day, $z=0$,   deceleration parameter  for the model is estimated to be  $q_0=-1.0109$; while $z_t=0.7516$ is  the best  estimate  acceleration-deceleration transition redshift value. 

 We  notice that the surface tension model behaves  like the $\Lambda$CDM  model at early time, while at late time (low redshift), the surface tension model acceleration is more significant than the standard cosmological model.

\section{Discussion}\label{Dis}
In this paper, we propose and analyze an alternative approach to explain the cosmic acceleration,  departing from a physical motivation. It was  shown  that the inclusion of a  surface tension  term can explain why the space-time fabric is  stretching out, and  leads  to modify the usual Friedmann equations.

The strength of the model resides in its simplicity. Due to the way the model was   developed, 
it preserves all the desirable  futures of the $\Lambda$CDM model, with one less parameter to be fixed.  It also possesses   desirable  futures of the phantom $w=-4/3$ cosmological model, without violating the null energy condition.

The model ensures  homogeneity and isotropy over the whole space-time - it explains: a) the Cosmological Coincidence Problem, b) why there is no need for such a cosmological constant to explain the acceleration of the Universe, and  c) why  we have not found any particle or fluid responsible for the dark energy component.  
Further more, it alleviates the tension between the discrepancy of the Hubble constant at late and early time epochs, concluding that the discrepancy on measurements of the Hubble constant  relies on the employed model.

We conclude that  the proposed model is in excellent agreement with observations.  The model  could be further developed to include additional  degrees of freedom, which  could be done by adding extra fluids or  considering a non flat-space cosmology. 

The aim of the paper is to introduce the implications of the  surface tension hypothesis  at large scales,   a more profound statistical analysis  would be presented elsewhere.

\section*{Acknowledgments}
Acknowledge the support provided by UAZ, project UAZ-2018-37554. \\



\bibliography{epjc}

\end{document}